\documentclass[aps,prl,superscriptaddress,twocolumn]{revtex4-1}
\usepackage{graphicx}                   
\usepackage{hyperref}                   
\usepackage{amsmath,amssymb}            
\usepackage{multirow}		
\usepackage{xcolor}	
\usepackage{amsmath,color}

\newcommand{\ave}[1]{\langle #1 \rangle}

\begin{document}

\author{Hor Dashti-Naserabadi}\email{hdashti@kias.re.kr}
\affiliation{School of Physics, Korea Institute for Advanced Study, Seoul 02455, Korea}
\author{Abbas Ali Saberi}\email{(corresponding author) ab.saberi@ut.ac.ir}
\affiliation{Department of Physics, University of Tehran, P. O. Box 14395-547, Tehran, Iran}
\affiliation{Institut f\"ur Theoretische
  Physik, Universit\"at zu K\"oln, Z\"ulpicher Str. 77, 50937 K\"oln,
  Germany}
\author{S. H. E. Rahbari}
\affiliation{School of Physics, Korea Institute for Advanced Study, Seoul 02455, Korea}
\author{Hyunggyu Park}
\affiliation{School of Physics, Korea Institute for Advanced Study, Seoul 02455, Korea}

\title{Two-dimensional Super-roughening in Three-dimensional Ising Model}


\begin{abstract}
We present a random-interface representation of the three-dimensional (3D) Ising model based on thermal fluctuations of a uniquely defined geometric spin cluster in the 3D model and its 2D cross section. Extensive simulations have been carried out to measure the global interfacial width as a function of temperature for different lattice sizes which is shown to signal the criticality of the model at $T_c$ by forming a size-independent cusp in 3D, along with an emergent super-roughening at its 2D cross section. We find that the super-rough state is accompanied by an intrinsic anomalous scaling behavior in the local properties characterized by a set of geometric exponents which are the same as those for a pure 2D Ising model. 
\end{abstract}

\maketitle


The microscopic definition of the surface of separation between two phases in the equilibrium systems and their transition from a smooth to a rough interface---the so-called roughening transition (RT)--- are among the long-standing problems in statistical physics \cite{burton1949crystal, burton1951, van1972phase, dobrushin1973gibbs, gallavotti1972phase, weeks1973structural, dobrushin1973gibbs, vanBeijeren1975, chui1976phase, swendsen1977monte, beijeren1977bcsos, burkner1983, mon19883dIsing, *Mon1988Erratum, landau1990rt3d, Hasenbusch1996, muller2005}.
The concept of RT in the context of crystal growth and its correspondence with the Ising model was first introduced by Burton and Cabrera \cite{burton1949crystal}. In this method, i.e. the lattice-gas realization of  the Ising model, the occupied sites corresponding to atoms are represented by spin up and vacancies are represented by spins down. In this picture, an interface separates the occupied sites from the rest of the system. It has been argued that there exists a temperature $T_R$ where the width of this interface diverges.

Let us briefly summarize the previous efforts in this regard during the past decades. Burton \textit{et al.} reported \cite{burton1951} that a RT occurs in the three-dimensional (3D) Ising model at a temperature $T_R $ very close to the critical point $T_c^{\mathrm{2D}}$ of a 2D Ising model, i.e., at $T_R \approx T_c^{\mathrm{2D}}\simeq 0.503 T_c$, with $T_c$ being the Curie point of the 3D Ising model. The arguments for the existence of such RT were based on mapping the interface problem into a 2D Ising model. This mapping is valid only at sufficiently low temperatures \cite{swendsen1977monte}. Dobrushin demonstrated that the interface width remains finite for low nonzero temperatures \cite{dobrushin1973gibbs}. Moreover, at low enough temperatures a sharp interface between areas of opposite magnetization exists.
From a different point of view, van Beijeren and Gallavotti \cite{van1972phase, gallavotti1972phase} have proved that there is no sharp interface for the 2D Ising model on a square lattice. They demonstrated that large fluctuations cause the interface width to diverge at any temperature even at very low nonzero $T$. Furthermore, they conjectured that the surface of separation between two phases of opposite magnetization in the 3D Ising model might show a RT. 
Weeks \textit{et al.} performed a low temperature expansion of the moments of the gradient of the density profile and used the slope at its midpoint to estimate the RT temperature $T_R$ for the width of an (001) interface in a 3D Ising model on a simple cubic lattice with isotropic and anisotropic coupling constants \cite{weeks1973structural}. In the case of anisotropic coupling constants, the so-called solid-on-solid (SOS) model, the vertical coupling constant $ J_z $  goes to infinity while the horizontal constants are fixed and finite $ J_x=J_y=J $. Moreover, they obtained a roughening temperature at $ T_R \approx 0.57 T_C $. 
van Beijeren proved a rigorous lower bound of the roughening point $ T_R \geq T_C^{\mathrm{2D}} $ for an arbitrary $ J_z $ \cite{vanBeijeren1975}.\\
\begin{figure}[t]
	\centering
	\includegraphics[width=0.48\textwidth]{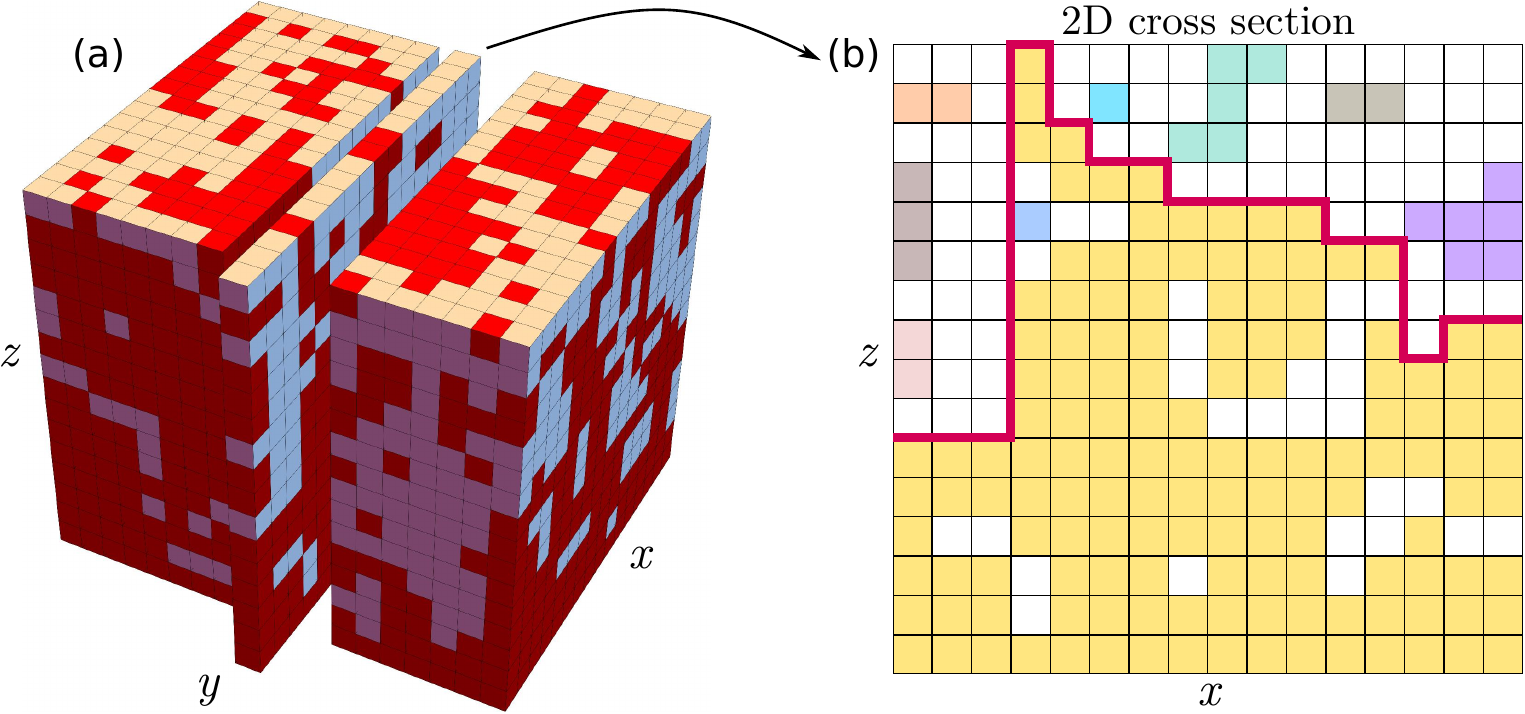}
	\caption{(a) Schematic illustration of geometric spin clusters in a 3D Ising model with fixed boundary condition at the bottom ($z$=0). (b) A 2D cross section of the model with its different spin clusters shown in different colors (spins up are merely colored). Note that the clustering procedure is performed independently in 3D and 2D on the same spin configuration. The solid line shows the unique interface on the 2D cross section which exhibits a super-roughening transition at the Curie point $T_c$. }
	\label{fig:height_def}
\end{figure}
Various Monte Carlo simulations have been also carried out on the 3D Ising model to clarify the RT problem \cite{swendsen1977monte, burkner1983, mon19883dIsing, landau1990rt3d, Hasenbusch1996}. Mon \textit{et al.} have done extensive simulations and determined the roughening temperature to be at $\sim 0.542(5)T_c$ \cite{landau1990rt3d}. They also found that for higher temperatures, the squared interface width increases logarithmically with system size.
Swendsen used Monte Carlo simulation to demonstrate the existence of the RT in SOS and discrete Gaussian (DG) models \cite{swendsen1977monte}. He described the relationship between the RT in SOS and DG models with phase transition in the 2D Ising model. In SOS models, interface overhangs and bubbles are neglected. A particular body-centered cubic SOS (BCSOS) model was introduced and solved exactly by van Beijeren \cite{beijeren1977bcsos}.
For precise simulation results of this and other models of the RT, see \cite{Hasenbusch1996}. We would like to emphasize that the RT does not
correspond to a bulk fixed point of the renormalization- group, and
studies of the RT have not led to progress in understanding the
critical behavior of the 3D Ising model.

\begin{figure}
	\centering
	\includegraphics[width=0.45\textwidth]{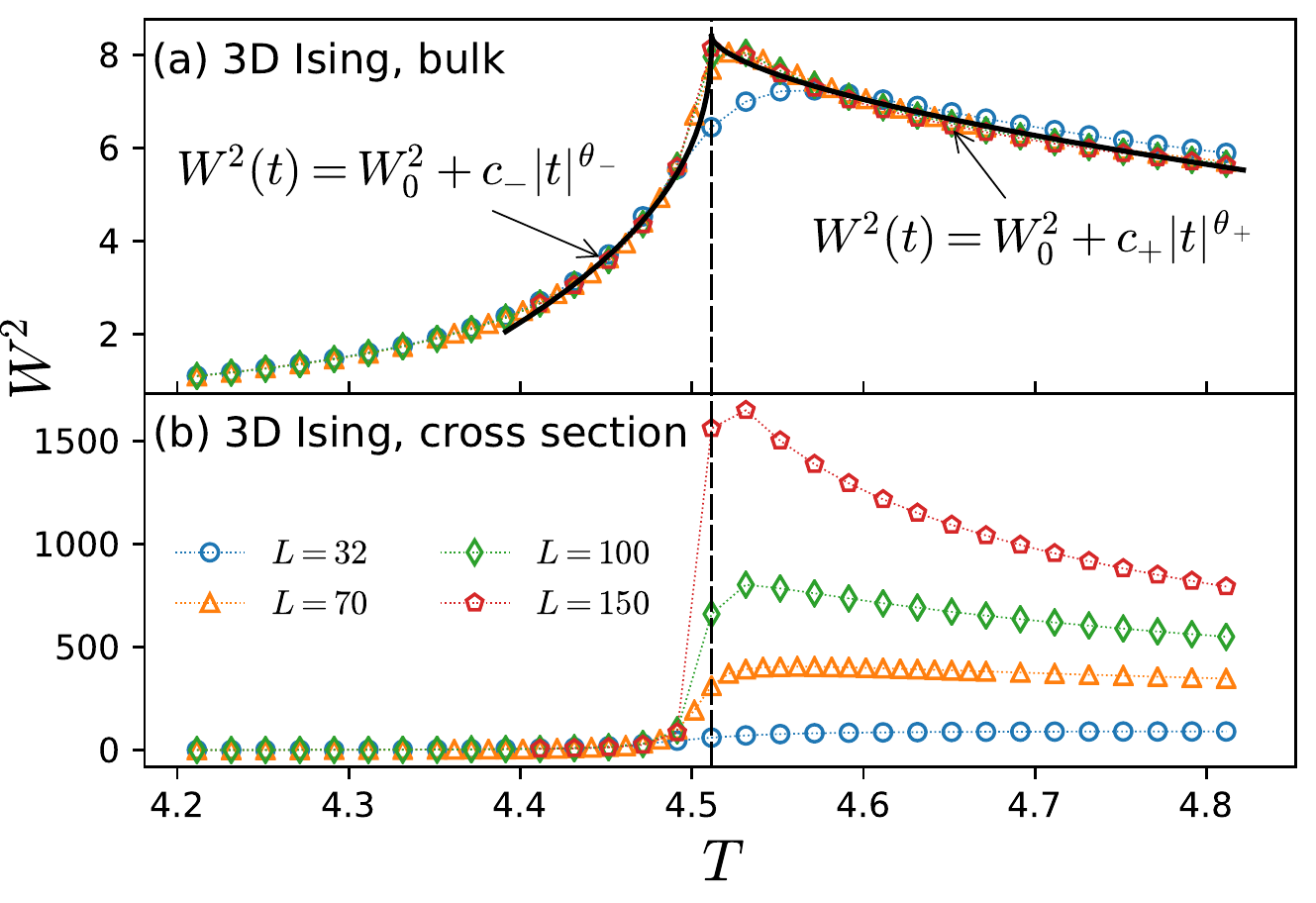}
	\caption{The average squared width as a function of temperature for (a) the fluctuating membranes in a 3D Ising model and, (b) the fluctuating curves at its 2D cross section. The vertical dashed line indicates the position of the Curie point best estimated numerically $T_c\approx 4.511524$ (all temperatures are expressed in [$J/k_B$] units) \cite{blote1995ising3d, blote1999cluster}. The solid lines show the scaling behavior of the cusp-like width in terms of the reduced temperature $t=(T-T_c)/T_c$ by introducing the supercritical $\theta_{+}=0.60(3)$ and subcritical $\theta_{-}=0.43(4)$ exponents near the critical point $t_c=0$. All averages are taken over independent spin configurations after thermal equilibration with $10^6$, $2.5\times 10^6$ and $5\times 10^6$ independent realizations for $T<T_c$, $T=T_c$ and $T>T_c$, respectively.  }
	\label{fig:WT}
\end{figure}

Here, we present an alternative approach to this problem by introducing some geometric measures in terms of thermal evolution of the spin domains' interface that exhibits a RT exactly at the Curie point $T_c $. We simulate the 3D Ising model by using the Wolff's single-cluster update algorithm \cite{wolff1989} on a cubic lattice of linear size $L$ whose spins at the bottom boundary ($z$=0) are set to be fixed at a state, say 'up'. Periodic boundary conditions along $x$ and $y$ directions and, free boundary condition at the top boundary are applied (Fig. \ref{fig:height_def}(a)). We focus on interfacial evolution of a uniquely defined  cluster of spins that is connected to the bottom boundary. A geometric spin cluster is defined as a set of connected nearest neighbor sites of like-sign spins which is identified by the Hoshen-Kopelman algorithm \cite{hoshen1976percolation}. With the interface we mean a random surface that separates the cluster attached to the floor from the rest of the spins. Such surface in the 3D system is a fluctuating membrane and in a 2D cross section of the system is a fluctuating curve (solid red line in Fig. \ref{fig:height_def}(b), which are the main subjects of the present study. For every identified random membrane (in 3D) and random curve (at the 2D cross section of the 3D model) we assign a unique corresponding height profile represented by $h(x,y)$ and $h(x)$, respectively, which are independent of each other since the clustering procedure is performed independently in the 3D and 2D crosse section on the same spin configuration (Fig. \ref{fig:height_def}). At every lattice point $\textbf{x}$ sitting at the floor (either at the floor of the 3D model denoted by $(x,y)$ or the 2D cross section of the model denoted by $(x,L/2) $), $h(\textbf{x})$ denotes the height of the uppermost spin which belongs to the cluster attached to the floor. This representation provides a (non one-to-one) map from spin configurations to a height profile. The unique feature of our approach is that it provides a representation of the 3D Ising model in lower dimension that signals the criticality of the bulk and also, it reveals unexpected similarities with the 2D Ising model at the Curie point. It is worth mentioning that our results are independent of the position of the 2D slice and it can be considered at any $1\le y\le L$ or $1\le x\le L$. 

\begin{figure}
	\centering
	\includegraphics[width=0.42\textwidth]{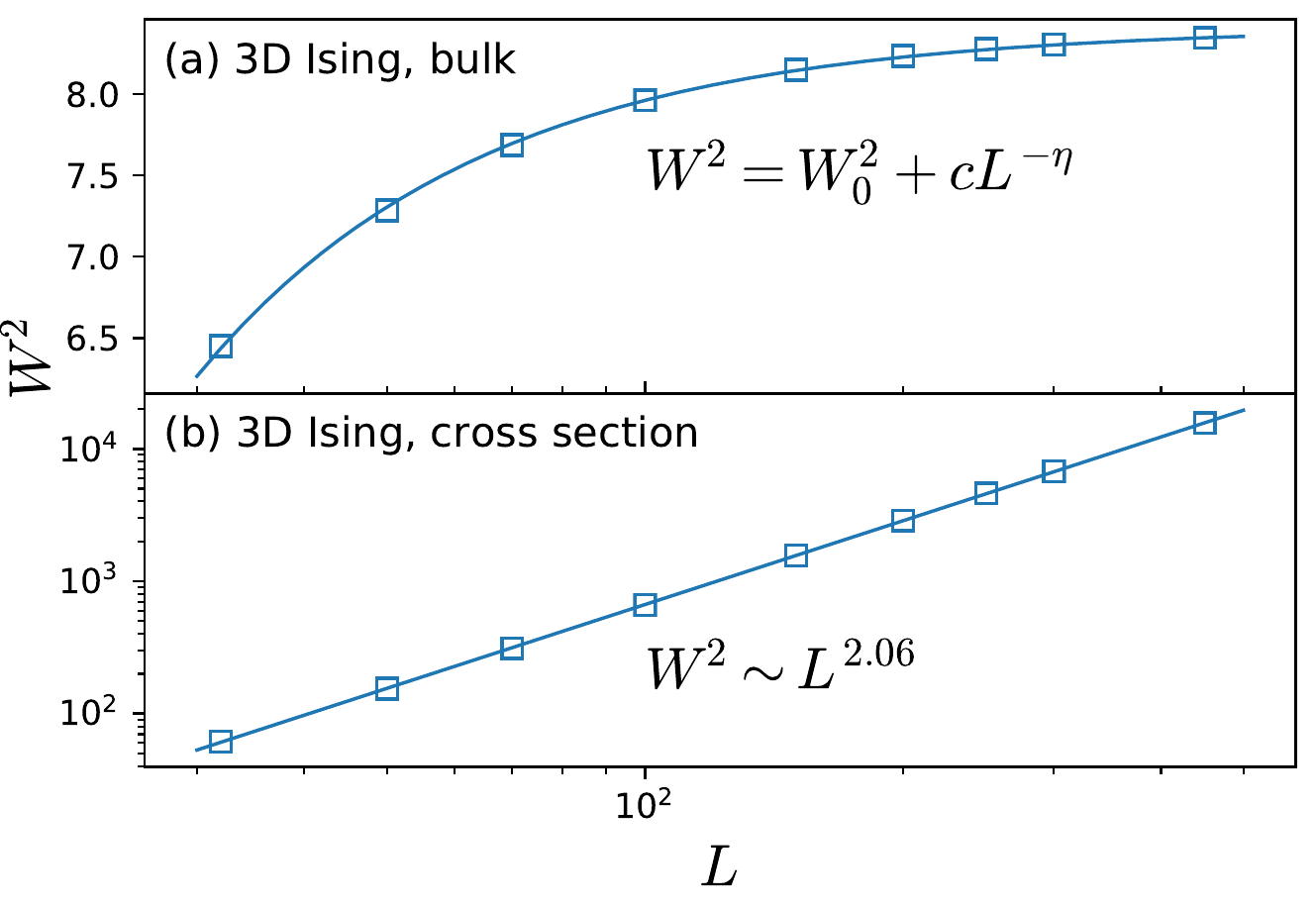}
	\caption{(a) The average squared width as a function of the system size $L$ at the Curie point $T=T_c$ for the fluctuating membranes in a 3D Ising model with the intrinsic roughness $W_0^2= 8.40(1)$, the constant $c = -208(9)$, and the irrelevant exponent $\eta = 1.33(2)$ estimated from the best fit to our data. (b) The same quantity for the fluctuating curves at the 2D cross section of the model at $T_c$ which diverges with the global roughness exponent $\alpha=1.03(2)$.}
	\label{fig:WL}
\end{figure}

Our aim here is to study the statistics of fluctuations in the height profiles $\{h(\textbf{x})\}$ within the proposed random-interface representation of the 3D Ising model. A basic quantity to characterize the height fluctuations around the mean value $\bar{h}$ is the global interface width, $W^2(L) = \ave{\overline{h^2(\textbf{x})}}_c:=\ave{\overline{[h(\textbf{x})-\bar{h}]^2}} $, where the bar stands for the average over all spatial space $ \textbf{x} $, and the brackets denote ensemble averaging. Previously posed definition of the interface by other authors are different (see Supplementary Information).
Figure \ref{fig:WT} presents the results of our computations for global width for different lattice sizes as a function of temperature for the fluctuating membranes (in the 3D Ising model, Fig. \ref{fig:WT}(a)) and the fluctuating curves (at the 2D cross section of the model, Fig. \ref{fig:WT}(b)). We find that the data for the 3D case are mostly coinciding for different system sizes. The only deviation is around the critical point $T_c$.  Interestingly, the width behavior signals the criticality of the bulk by forming a cusp exactly at $T=T_c$ (Fig. \ref{fig:WT}).
To further investigate the system size effects at $T_c$, we have produced the data for global width at $T_c$ for larger number of sizes and examined if it exhibits a scaling behavior. As shown in Fig. \ref{fig:WL}(a), the best fit to our data suggests the relation
\begin{equation}
W^2(L)=W_0^2+c L^{-\eta},
\end{equation} 
with the irrelevant exponent $\eta=1.33(2)$, the constant $c=-208(9)$, and the intrinsic size-independent surface width $W_0^2=8.40(1)$. 
 We also find that in terms of the reduced temperature $t \equiv (T-T_c)/T_c$, the global width follows the scaling relation 
\begin{equation}
W^2(t) = W_0^2 + c_{\pm}|t|^{\theta_{\pm}}, 
\end{equation}	
with  $c_{+}=-14.6(12), \theta_{+}=0.60(3)$ for $t>0$ and $c_{-}=-30(4), \theta_{-}=0.43(4)$ for $t<0$ near the critical point $t_c=0$ (Fig. \ref{fig:WT}(a)).
The surprise comes from the fact that the percolation transition of spin clusters (as a pure geometric transition) occurs at some temperature $T_p\sim 4.31$ \cite{muller1974droplet, saberi2010ising3d} well below the Curie point $T_c\sim 4.51$, and one would naturally expect that $W^2_0$, as a geometric quantity, should respond to the global geometric changes at $T_p$, but it doesn't, and, in turn, it signals the thermal phase transition in the 3D Ising model.

The intrinsic width characterizes the internal structure of the fluctuating membrane which is due to the holes and overhangs mostly dominant at $T_c$ for which the leading contribution comes from the short wavelength fluctuations in the local height increments. This behavior is totally different from that of the rough surfaces \cite{barabasi1995fractal, family1985scaling, krug1997origins} for which the Family-Vicsek scaling ansatz, i.e.,  $W^2(L) \sim L^{2\alpha}$, holds at the steady state where $\alpha>0$ is the global roughness exponent, originating from the long-wavelength fluctuations. Existence of such small length scale at the critical point may explain why in contrary to the 2D Ising model, geometric spin clusters do not capture the scale-invariant criticality of the 3D model in a way that the Fortuin-Kasteleyn clusters do \cite{fortuin1972random}. However, the quantity $W^2(T)$, built on the geometric spin clusters,
is able to capture the criticality by forming a cusp at  $T_c$.

\begin{figure}[t]
	\centering
	\includegraphics[width=0.43\textwidth]{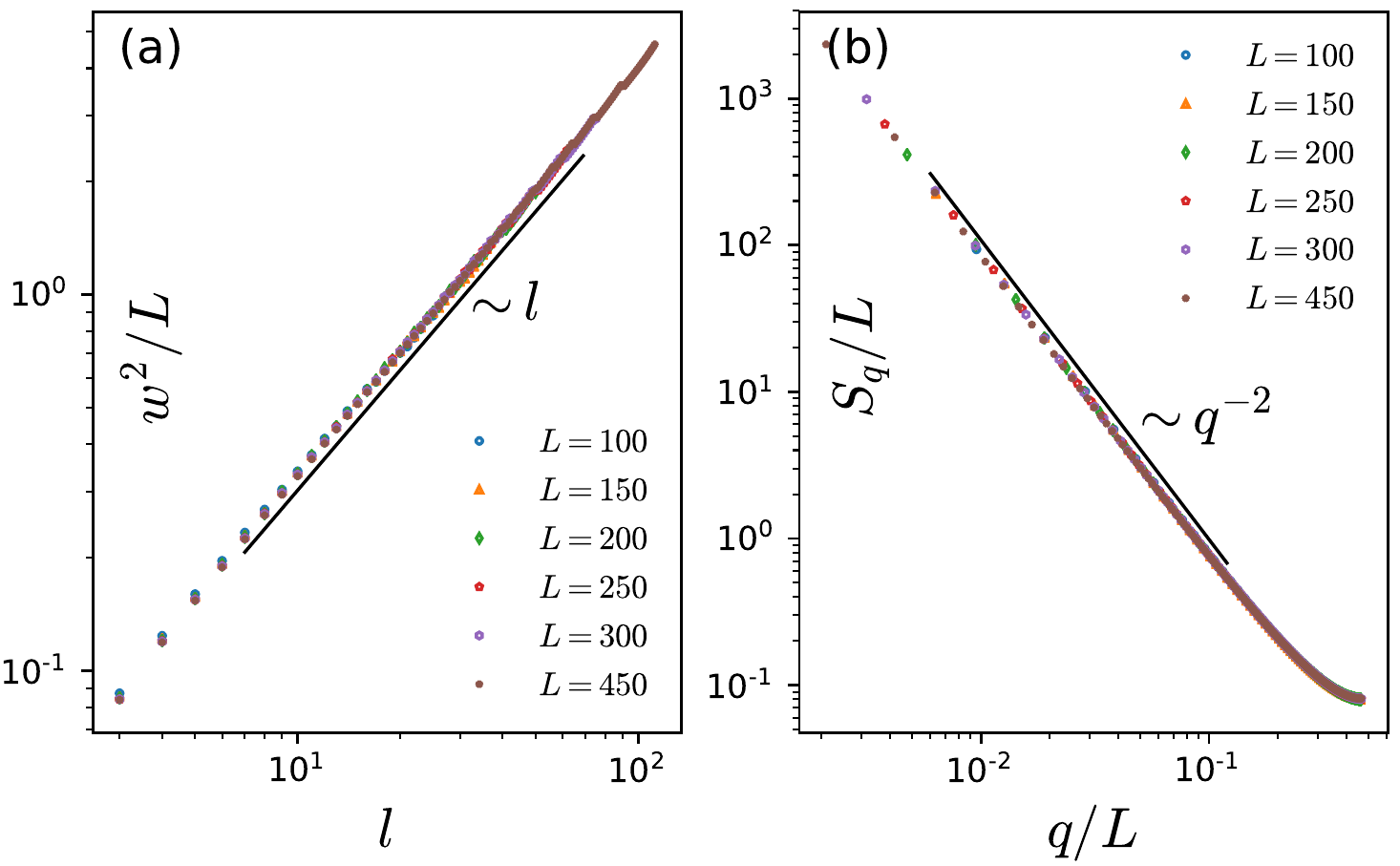}
	\caption{Scaling behavior of two local measures computed on the 2D cross section of the 3D Ising model of various linear size $L$ at the Curie point $T_c$. (a) Scaled squared local width $w^2(l,L)$ as a function of the window size $l$ and, (b) the power spectrum $S_q(L)$. Their anomalous scaling properties give two corresponding local exponents $\alpha_l=\alpha_s\approx0.5$ different from the global roughness exponent $\alpha\approx1$ (Fig. \ref{fig:WL}(b)).}
	\label{fig:wl_sq}
\end{figure}

In order to show that the emergence of the intrinsic width is a characteristic feature of the three dimensions, let us now look at the statistics of the height profile built on a 2D cross section of the spin configuration at $y=L/2$ (Fig. \ref{fig:height_def}) in the 3D Ising model. Surprisingly, the global interface width exhibits a totally different behavior at the 2D cross section of the model with a geometric RT (Fig.~\ref{fig:WT}(b)). For $T<T_c$
the interface width remains small as $L$ increases, indicative of a smooth interface in the sub-critical regime, while it is non-zero in the super critical region with $T>T_c$. Exactly at the critical point $T=T_c$, the global interface width diverges with the system size, i.e., $W^2(L)\sim L^{2\alpha}$ with the global roughness exponent $\alpha=1.03(2)$ (Fig. \ref{fig:WL}(b)) which is a super-rough interface. The global roughness exponent $\alpha\sim 1$ guaranties the fractal property of the interfaces \cite{Mandelbrot1977} (i.e., the fluctuating curves) but it strongly suggests the existence of an anomalous scaling behavior implying that one more exponent, i.e., the local roughness exponent $\alpha_{l}$, may be needed to assess the universality class of the model.

\begin{figure}[b]
	\centering
	\includegraphics[width=0.45\textwidth]{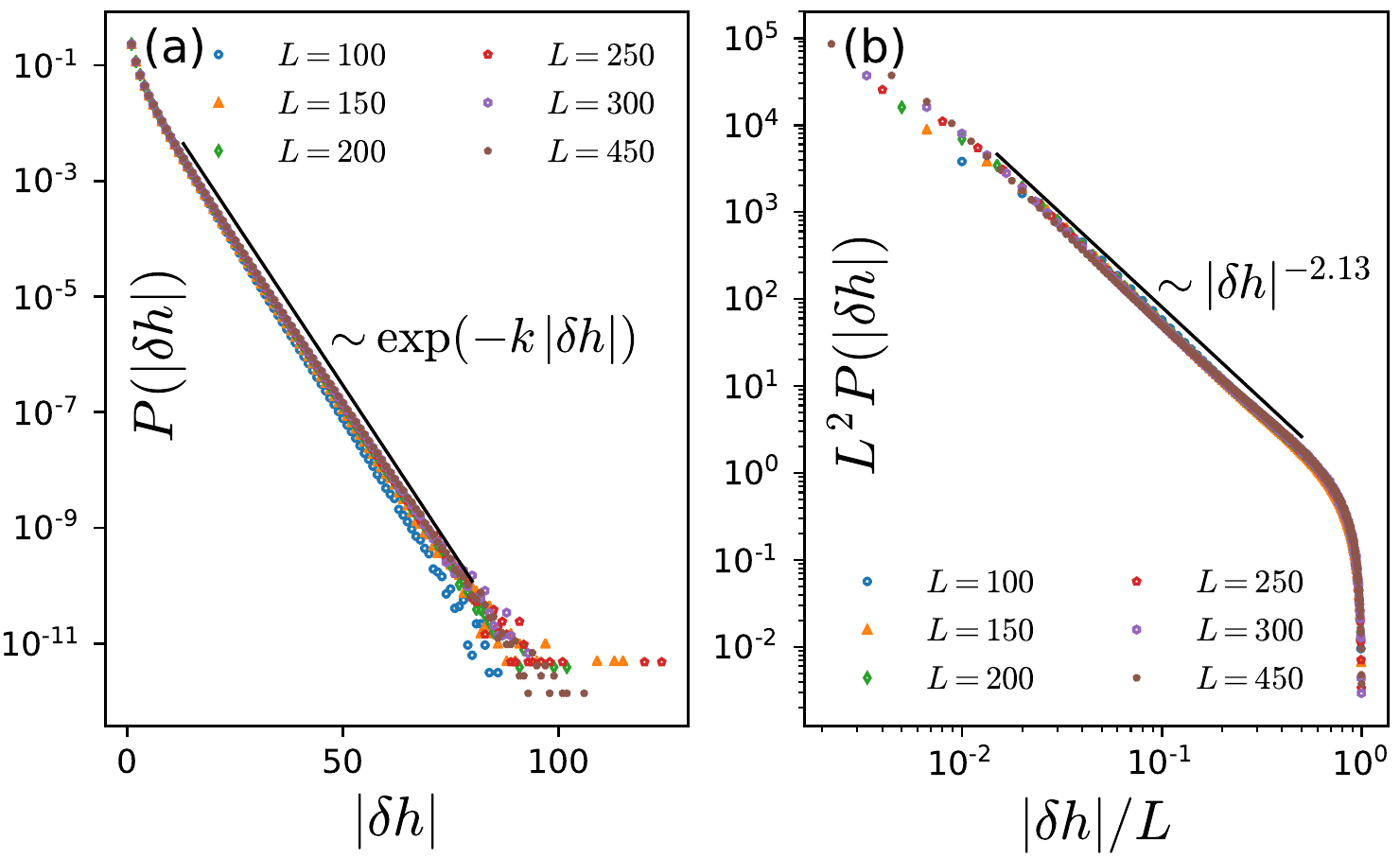}
	\caption{Probability distribution function of the absolute height differences $|\delta h|$ in the 3D Ising model (a) and its 2D cross section (b) for various linear size $L$ at the Curie point $T_c$. Our data is consistent with a size-independent exponential distribution of the height fluctuations in 3D (with $k=0.26$), and a power-law distribution $\sim |\delta h|^{-\tau}$ with $\tau\approx2.13(2)$ in a 2D cross section of the model in which the rescaled data for different sizes collapse onto a single curve.} 
	\label{fig:pdf_dh_3D}
\end{figure}

In order to examine this anomalous scaling  hypothesis, let us investigate the scaling behavior of the two following local measures at $T=T_c$: (i) The local interface width $w^2(l) := \pmb{\left\langle\vphantom{\frac{1}{2}}\right.}
\langle [h(x)-\ave{h}_l]^2 \rangle_l
\pmb{\left.\vphantom{\frac{1}{2}}\right\rangle}$, where $ \langle \cdots\rangle_l $ indicates an average over $ x $ in windows of size $ l $ that is expected to have the scaling relation $ w^2(l) \sim l ^{2\alpha_l} $, with $ \alpha_l $ being the local roughness exponent \cite{barabasi1995fractal}. The extra bold brackets denote for the ensemble averaging. (ii) The structure factor (or the power spectrum) $ S_q = \langle \hat{h}(q)\hat{h}(-q) \rangle $, in which the Fourier transform of the height profile $h(x)$ is given by $\hat{h}(q) = L^{-1/2} \sum [h(x) - \bar{h}] \exp(iqx)$, which is supposed to follow the power law $ S_q \sim q^{-(2\alpha_s+1)} $ \cite{barabasi1995fractal}, with the spectral roughness exponent $\alpha_s$. The relation $\alpha_l = \alpha_s=\alpha$ is only valid for the self-affine surfaces that follow the Family-Vicsek scaling as one of the possible scaling forms compatible with generic scaling invariant growth \cite{superroughening1997, ramasco2000generic, scaling2005lopez}, which is not the case here. Figure~\ref{fig:wl_sq} represents the results of our computations for the local width (\ref{fig:wl_sq}(a)) and the power spectrum (\ref{fig:wl_sq}(b)) for an ensemble of interfaces on a 2D cross section of the 3D Ising model at $T=T_c$ for various system size $L$. We find that all data for different size $L$ collapse onto a single curve  when they are suitably rescaled, and they follow the scaling relations $ w^2(l,L) \sim  l^{2\alpha_l} L^\alpha $ and $ S(q,L) \sim q^{-(2\alpha_s+1)} L^{2(\alpha-\alpha_s)}$, respectively, with $\alpha\approx 1$ again, and $\alpha_l=\alpha_s\approx 0.5$ estimated from the best fit to our data. Apparently these exponents do not belong to the Family-Vicsek scaling, however, within the generic scaling picture presented in \cite{ ramasco2000generic}, they fall into the class of the intrinsically anomalous roughened surfaces.

\begin{table}
	\caption{Scaling exponents related to the random interfaces of the 2D cross section of the 3D Ising model and the 2D Ising model.}
	\begin{tabular}{|c|c|c|}
		\hline 
		exponent & cross section of 3D Ising  & 2D Ising \\ 
		\hline 
		$ \alpha $ & $1.03(2)$ & $1.00(1)$  \\ 
		\hline 
		$ \alpha_l $ & $0.52(2)$ & $0.51(1)$   \\ 
		\hline 
		$ \alpha_s $ & $0.50(1)$ & $0.51(1)$   \\ 
		\hline 
		$ \tau $ & $2.13(2)$ & $2.15(2)$ \\ 
		\hline 
	\end{tabular} 
	\label{tab:exp}
\end{table}

\begin{figure}
	\includegraphics[width=0.45\textwidth]{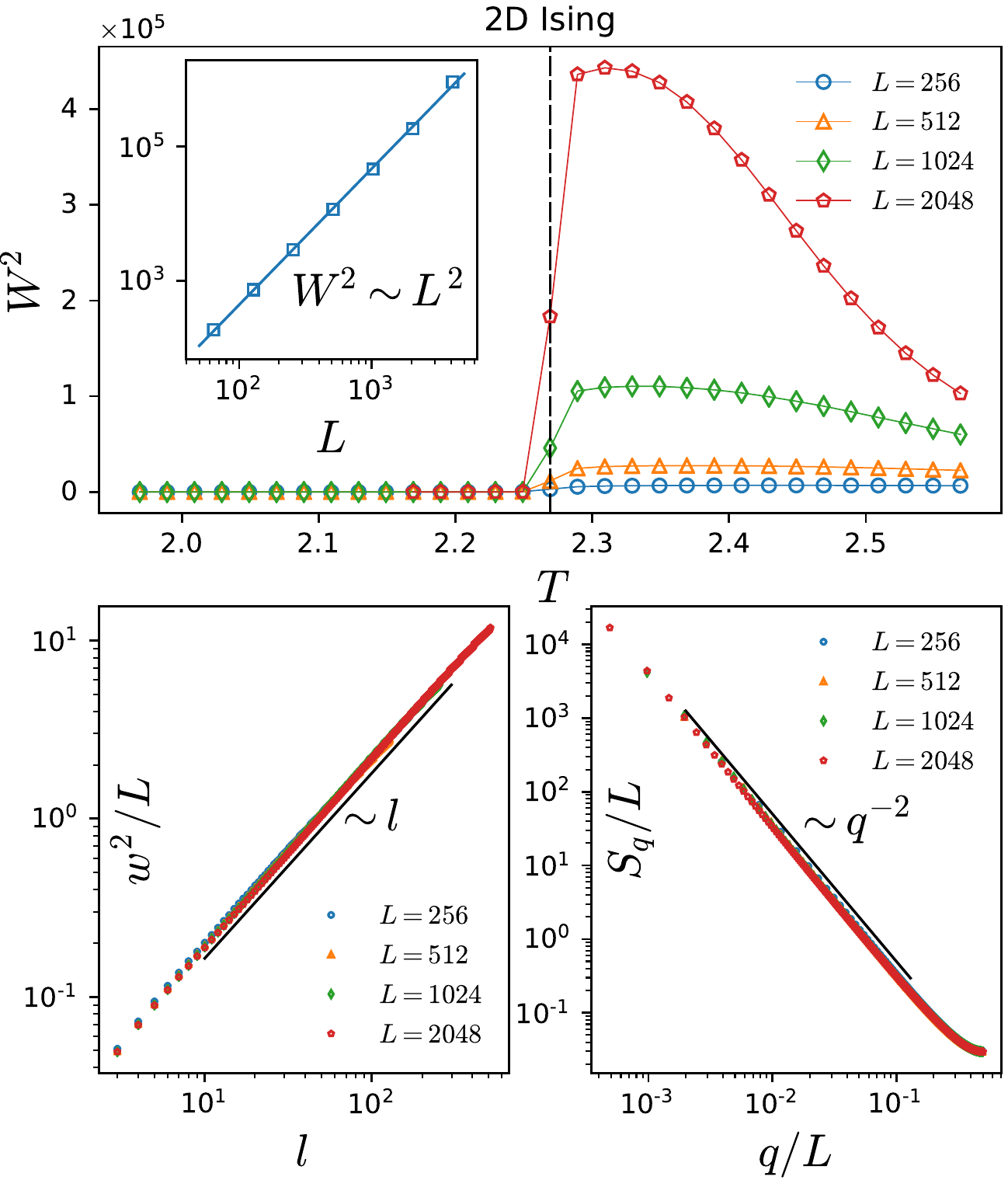}
	\caption{
		Upper panel: The average squared width as a function of temperature for the fluctuating curves in a 2D Ising model. The vertical dashed line indicates the position of the Curie point, $T_c^{\mathrm{2D}} = 2/(\ln(1+\sqrt{2}))$. Inset: The average squared width as a function of the system size $ L $ at $ T = T_c^{\mathrm{2D}} $. The global width $ W $ diverges with the global roughness exponent $ \alpha = 1.00(1) $. lower panels: Scaling behavior of two local measures computed on the 2D Ising model of various linear size $L$ at the critical point $T=T_c^{\mathrm{2D}}$.  Scaled squared local width $w^2(l,L)$ as a function of the window size $l$ (left) and, the power spectrum $S_q(L)$ (right). Their scaling properties give two corresponding local exponents $\alpha_l = \alpha_s \approx 0.5$.
	}
	\label{fig:2d}
\end{figure}

The statistical measures discussed here for the random-interface representations of the 3D Ising model are governed by the properties of the corresponding height fluctuations which can be characterized by the probability distribution of the height differences $\delta h$ between any pair of nearest neighbor sites. As Fig.~\ref{fig:pdf_dh_3D}(a) shows, the distribution of the height fluctuations in the random-membrane representation of the 3D Ising model at $T_c$ is a size-independent exponential, i.e., $P(|\delta h|,L)\sim \exp(-k |\delta h|)$ with $k=0.26$. This may explain why the membrane in 3D is smooth due to the exponential suppression of large fluctuations. Emergence of the size-independent intrinsic width $W_0$, is also in connection with the observed size-independent distribution, since the exponential distribution naturally introduces a finite length scale $\propto 1/k$ in the system. We find that the height fluctuations in the random-curve representation of a 2D cross section of the 3D model behave totally different and follow a scaling distribution  $ P(|\delta h|, L) \sim |\delta h|^{-\tau} L^{-(2+\tau)}$ with $ \tau \approx 2.13(2)$. This power-law distribution is strongly consistent with the previous observation by two of us in \cite{saberi2010ising3d} that the geometric spin clusters in the 2D cross section become critical exactly at the critical point of the 3D bulk. To give more evidence on the critical manifestation of the 2D cross section, we studied the random-curve representation of the pure 2D Ising model at criticality and, interestingly, found the same results as for the 2D cross section of the 3D Ising model with the conjectured super-universal exponents $\alpha=1$, and $\alpha_l=\alpha_s=1/2$. 
Table \ref{tab:exp} summarizes the global and local exponents that we have obtained for the 2D Ising model and the cross-section of the 3D Ising model.

As shown in Fig. \ref{fig:2d} (upper panel), the global interface width for the 2D Ising model exhibits a geometric roughening transition at $T= T_c^{\mathrm{2D}}$. This behavior is very similar to the one observed on the cross section of the 3D Ising model (see Fig. \ref{fig:WT} (b)). This similarity is also supported by computing the local measures addressed in Fig. \ref{fig:2d} (lower panels). This figure represents the results for the local width $w(l)$ and the power spectrum $S_q$ for an ensemble of interfaces of the 2D Ising model at $T_c^{\mathrm{2D}}$ for various system size $ L $. We find the local roughness exponent and the spectral exponent as $\alpha_l = \alpha_s \approx 0.5$ (see Table \ref{tab:exp}).

\textit{Acknowledgment.} A.A.S. would like to acknowledge the supports from the Alexander von Humboldt Foundation and the research council of the University of Tehran. We would like to thank the High Performance Computing (HPC) center in the University of Cologne, Germany, where the most of computations have been carried out.  We also thank the KIAS Center for Advanced Computation for providing computing resources.
	
\bibliography{refs}

\end{document}